# Complex nature of magnetic field-induced ferroelectricity in GdCrTiO$_5$

T. Basu[1,2,*,#], D. T. Adroja[3,4], F. Kolb[1], H.-A. Krug von Nidda[1], A. Ruff[1], M. Hemmida[1], A. D. Hillier[3], M. Telling[3], E.V. Sampathkumaran[2], A. Loidl[1] and S. Krohns[1]

[1]*Experimental Physics V, Center for Electronic Correlations and Magnetism, University of Augsburg, Universitätsstrasse 2, D-86135 Augsburg, Germany*
[2]*Tata Institute of Fundamental Research, Homi Bhabha Road, Colaba, Mumbai-400005, India*
[3]*ISIS Facility, Rutherford Appleton Laboratory, Chilton, Didcot Oxon, OX11 0QX, United Kingdom*
[4]*Highly Correlated Matter Research Group, Physics Department, University of Johannesburg, PO Box 524, Auckland Park 2006, South Africa*

*tathamaybasu@gmail.com
[#] Present address: Laboratoire CRISMAT, UMR 6508 du CNRS et de l'Ensicaen, 6 Bd Marechal Juin, 14050 Caen, France.

## Abstract

This work shows an unconventional route for spin-driven ferroelectricity originating from a metastable magnetic field-induced canting of the chromium sublattice in the presence of gadolinium moments in GdCrTiO$_5$ at low temperatures. Compared to the isostructural neodymium compound, significant differences of magnetism and magnetoelectric effects are seen. We present the results of thorough investigations of temperature and magnetic field dependent magnetization as well as *ac* and *dc* magnetic susceptibility. These bulk measurements are complemented by local-probe spectroscopy utilizing electron-spin resonance and muon-spin rotation/relaxation for probing the chromium moments. Ferroelectric order is inferred from pyro- and magnetocurrent measurements. GdCrTiO$_5$ shows a pyrocurrent signal around 10 K, only if the system is cooled in an applied magnetic field exceeding 10 kOe. A distinct spin-driven ferroelectric order is revealed in this state for temperatures below 10 K, which can be switched by changing the magnetic-field direction and the polarity of the electric field. The magnetic measurements reveal no clear signature of long-range magnetic ordering. The presence of such 'meta-magnetoelectric-type' behaviour in the absence of any 'meta-magnetic' behaviour is rare in literature. Our microscopic spectroscopy results indicate significant changes of the magnetic properties around ~10 K. Probably there exists exchange frustration between Gd and Cr moments, which prevents long-range magnetic ordering at higher temperatures. Below 10 K, weak ferromagnetic magnetic order occurs by minimizing frustration due to lattice distortion, which supports magnetodielectric coupling. However, non-polar distortions attain appreciable values after application of magnetic fields above 10 kOe, obviously breaking spatial inversion symmetry and creating ferroelectricity.

1. **Introduction**

Multiferroic compounds with magnetoelectric coupling (cross-coupling between spin and dipolar degrees of freedom) have generated considerable interest in recent years due to a variety of magnetoelectric (ME) phenomena, interesting for basic research as well as for potential applications in device technology.[1,2] In non-collinear magnets, spin-driven ferroelectricity is induced via the inverse Dzyaloshinskii-Moriya (DM) interaction, and in collinear magnets either via exchange-striction or in some cases via local distortions, due to a spin-dependent p-d hybridization.[2,3,4] Nevertheless, there are many systems (such as, Haldane spin-chain systems), where the cross-coupling mechanism is not well understood.[5,6,7] Even in well-known multiferroic compounds like $R$Mn$_2$O$_5$ ($R$ = rare-earth),[8] crystallizing in an orthorhombic structure with *Pbnm* space group, the mechanism is still under debate.[9,10] Initially, there were contradictory reports on the mechanism of magnetoelectricity. Later the existence of both DM interaction and exchange-striction in these compounds has been demonstrated.[9] However, recently it has been shown by Balédent *et al.*[10] that, for this class of multiferroics, ferroelectricity is already present at room temperature, deep in the paramagnetic regime, and is further enhanced at low temperatures via spin-driven mechanisms.

Another family of compounds of type $R$CrTiO$_5$,[11,12] crystallizing in the same structure as $R$Mn$_2$O$_5$, has received considerable attention.[13] Early neutron diffraction experiments revealed long-range antiferromagnetic order in NdCrTiO$_5$ below 13 K.[11] However, magnetoelectric coupling and the onset of spin-driven ferroelectricity have been observed already at 21 K and were attributed to possible antiferromagnetic ordering of the Cr moments.[12,13] Hwang *et al.*[13] predicted that the Cr and Nd moments order around 21 K and 13 K, respectively. They proposed the possibility of magnetostriction mechanism for the ME coupling. Later on, Kori et al.[14] argued that both Nd and Cr moments start to order around the same temperature, which then allows for a DM interaction in a possible non-collinear spin structure.

Further investigations of $GdCrTiO_5$, a heavy rare-earth member of this series, by Basu et al,[15] evidenced a magnetic-field ($H$) induced dielectric anomaly around 10 K, similar to the observations in $NdCrTiO_5$. However, there was no clear-cut evidence of long-range magnetic order – neither in the temperature dependence of the magnetic susceptibility ($\chi$) nor in the heat capacity ($C$) – for $GdCrTiO_5$.[15] Unlike in $NdCrTiO_5$, the derivative of $\chi$ indicates possible magnetic ordering from Cr around 10 K, warranting confirmation by microscopic techniques. According to the de Gennes[16] scaling – i.e. the expected linear relation between magnetic-ordering temperature and de Gennes factor $dG = (g_J-1)^2 J(J+1)$, where $J$ denotes the total spin and $g_J$ the Landé factor of the rare-earth ion – a reduction of the magnetic ordering temperature is not typical for heavy rare-earth members. Therefore, finding out the possible mechanism and the role of different magnetic ions on the magnetic and magnetoelectric properties of this class of compounds is a challenging task. Until now, to the best of our knowledge, there exists no report using local spectroscopic probes on $RCrTiO_5$ compounds, focusing on these fundamental aspects of multiferroicity. Local probes are expected to shed some light on existing magnetic interactions in this class of materials and to unravel the role of rare earth and chromium moments on the ME coupling.

Here, we report the results of our investigations on $GdCrTiO_5$ by means of electron-spin-resonance (ESR) and muon-spin rotation/relaxation (μSR) spectroscopy. Neutron-scattering experiments are not feasible due to high absorbance of Gd, whereas both ESR and μSR spectroscopy are ideal tools to explore the microscopic spin-dynamics and magnetic correlations on a local scale. We also have investigated in detail the electric polarization in the presence of magnetic fields exploring the role of the rare-earth ions on ME properties. Our results reveal spin-driven ferroelectricity for temperatures below 10 K under the application of magnetic fields above 10 kOe. Although the features due to the onset of long-range magnetic order are not transparent in the magnetic susceptibility and heat capacity down to 2 K, ESR and μSR evidence sharp magnetic anomalies around 10 K, which support the existence of a magnetic phase transition at this temperature. We also discuss possible scenarios for magnetic interactions and ME properties in $RCrTiO_5$ compounds.

**2.     Experimental Details**

Preparation and structural characterization of polycrystalline $GdCrTiO_5$ is described in an earlier work.[15] The *dc* magnetization (*M*) as a function of temperature and magnetic field was carried out using a Superconducting Quantum Interference Device (SQUID), procured from Quantum Design. The same instrument was used to measure *ac* susceptibility χ. Electron spin resonance measurements were performed in a Bruker ELEXSYS 500 spectrometer with a standard rectangular microwave cavity ER 4102 ST working at X-band frequency (ν = 9.36 GHz). An Oxford ESR 900 helium gas-flow cryostat was used for the ESR measurements in the temperature range of 4.2 < *T* < 300 K. By sweeping the external static magnetic field in a regime 0 < *H* < 10 kOe at constant microwave frequency, ESR measures the microwave power absorbed from the transverse magnetic component of the microwave field in the center of the cavity. Due to the lock-in amplification with field modulation, the field derivative d*P*/d*H* of the absorption signal is detected as function of the external magnetic field. For the ESR measurements, the powder form of the sample was immersed in paraffin within a suprasil-quartz tube. The μSR experiments were carried out under zero-field (ZF) conditions down to 2 K on the HIFI spectrometer at the ISIS pulsed muon source of the Rutherford Appleton Laboratory (UK).[17] The powder sample was mounted on a silver plate with GE-varnish and cooled to a base temperature of 2 K in He-exchange gas cryostat. The 'MANTID'[18] software was used to analyse the data. Pyroelectric current measurements were carried out as function of magnetic field and temperature using an electrometer (6517A, Keithley) employed in a Physical Property Measurement System (PPMS, Quantum Design). The temperature-dependent remnant electric polarization ($P_r$) has been determined by cooling the sample in the presence of an electric field from 50 to 4 K. Subsequently, the electric field was set to zero, the capacitor shortened for sufficiently long time to remove any stray charges and then the pyroelectric current ($I_{pyro}$) was measured as a function of temperature with a constant heating rate. For pyroelectric current measurements in magnetic fields, two conditions, field-cooled (FC) and zero-field-cooled (ZFC), were carried out. The sample is cooled in the presence of an electric and a magnetic field for FC conditions. For ZFC condition no magnetic field is applied while cooling the sample. Subsequently, the pyroelectric current (FC and ZFC conditions) is measured while heating the sample in presence of applied magnetic fields. Magnetocurrent ($I_{magneto}$) measurements were performed at a fixed temperature as a funtion of magnetic fields.

### 3. Results
### A. Magnetic properties

The temperature dependences of the *dc* susceptibility $\chi$ (= *M/H*) and *C* of GdCrTiO$_5$ have been reported in Ref. 15. It may be recalled that no clear-cut signature due to magnetic ordering was observed, though the Curie-Weiss temperature ($\Theta$) was found to be approximately $\Theta$ = -25 K. The deviation from the high temperature linear inverse magnetic susceptibility appears already below 150 K indicating the presence of significant antiferromagnetic exchange (short-range interaction) in this compound.[15] In addition, the derivatives of $\chi(T)$ and $C(T)$ revealed a change of slope around 10 K providing possible evidence for the onset of weak magnetic order of the Cr moments.[15]

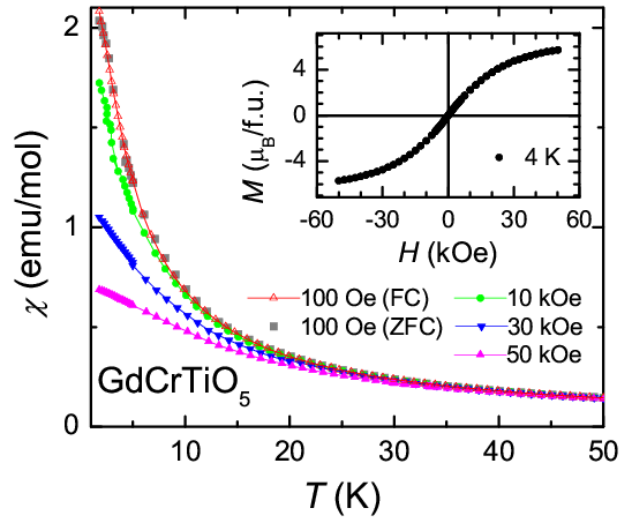

**Figure 1:** Temperature dependent magnetic *dc* susceptibility of GdCrTiO$_5$ for various applied magnetic fields between 100 Oe and 50 kOe. The zero-field cooled and field cooled data is shown only for 100 Oe. The inset shows the magnetization at 4 K for ZFC conditions.

Though the *dc* susceptibility $\chi(T)$ (= *M/H*) of the titled compound was already reported,[15] here, we will show, for the sake of completeness, the data measured in the course of the present experiments at low temperatures. Field-cooled measurements of the magnetic susceptibility in external magnetic fields between 100 Oe and 50 kOe are shown in Fig. 1 and compared with the zero-field cooled susceptibility measured with 100 Oe. FC and ZFC measurements are identical within experimental uncertainty and show no indications of a bifurcation, an observation that excludes any spin-glass type of freezing of magnetic moments. For all fields, the magnetic susceptibility shows a continuous increase with decreasing temperature, characteristic of purely paramagnetic behaviour. There is no indication of any anomaly pointing towards the existence of a magnetic phase transition. The inset in Fig. 1 shows the magnetization in fields from - 50 to + 50 kOe at 4 K. The field dependence of the magnetization is very close to an ideal Brillouin function, expected for purely paramagnetic behaviour.

The real part ($\chi'$) of the *ac* susceptibility as a function of temperature below 20 K, measured in the absence of any *dc* magnetic field, using a 2 Oe amplitude for frequencies between 1 and 1139 Hz, is presented in Fig. 2. Like the *dc* susceptibility, *ac* susceptibility also exhibits a continuous increase on decreasing temperature, suggestive of purely paramagnetic behaviour without any indication for an onset of long-range magnetic order or spin-glass freezing. The latter facts are further exemplified by the temperature dependence of the imaginary part of *ac* susceptibility ($\chi''$), shown for 1 Hz in the inset of Fig. 2. The *ac* loss is always close to zero with no indication for spin order.

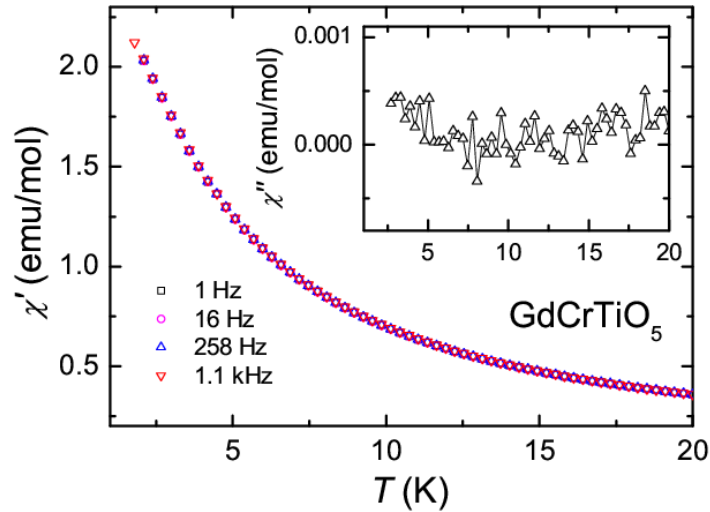

**Figure 2:** Real part of *ac* susceptibility of GdCrTiO$_5$ as a function of temperature for various frequencies in the absence of a *dc* magnetic field. Inset shows imaginary part of *ac* susceptibility for 1 Hz.

Figs. 1 and 2 are paramount examples of paramagnetic behaviour of a local moment system. However, one has to be aware that the Gd moments with spin $S = 7/2$ are much larger than the Cr moments with $S = 3/2$. Thus, Gd dominates the paramagnetic susceptibility and probably could hide effects due to (weak) magnetic ordering of Cr. To evidence possible ordering of the chromium moments, we therefore conducted experiments utilizing local probes, such as µSR and ESR.

## B. ESR Spectroscopy

Fig. 3 shows typical ESR spectra of GdCrTiO$_5$ at selected temperatures. One observes a single broad resonance line, which is well described by a Lorentzian curve with resonance field $H_{res}$ and a half-width at half-maximum (HWHM) linewidth $\Delta H$. Due to the large linewidth, which is of the same order of magnitude as the resonance field, it was necessary to take into account the mirrored resonance at negative resonance field $-H_{res}$ for fitting the line shape as described, e.g., by Joshi and Bhat.[19]

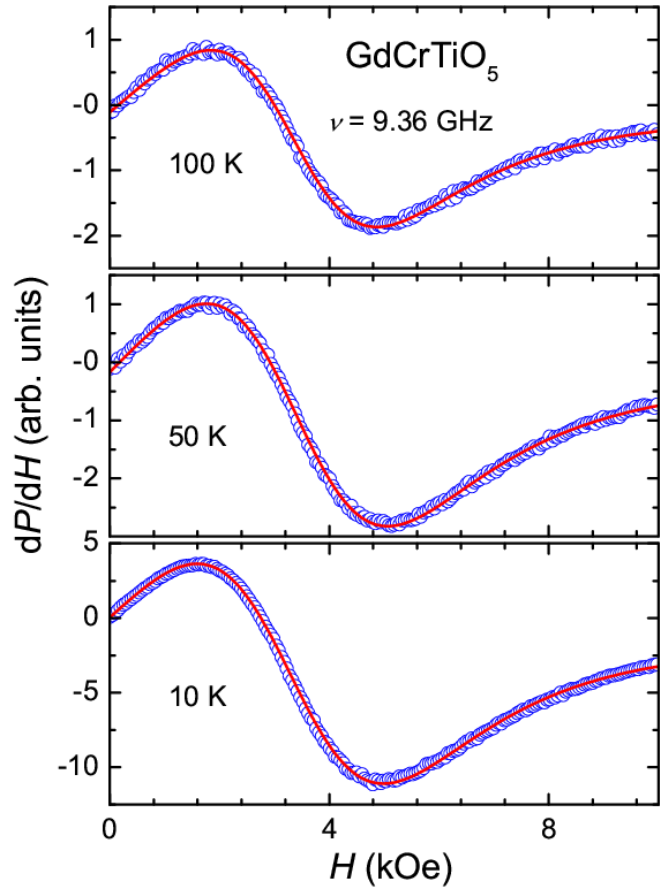

**Figure 3:** ESR spectra of GdCrTiO$_5$ at selected temperatures (100 K, 50 K and 10 K) as described in the text. Shown is the field derivative of the absorption $dP/dH$ vs. the external magnetic field $H$. The solid lines are fits as described in the text.

The resulting fit parameters are depicted in Fig. 4. The double integrated intensity $I_{ESR}$ (Fig. 4a), which is shown in the top frame, follows a Curie-Weiss law $1/I_{ESR} \sim T - \Theta$ at high temperatures, with an antiferromagnetic Curie-Weiss temperature of $\Theta \sim -25$ K, but deviates at lower temperatures. This is in reasonable agreement with the static susceptibility indicating that both the $Gd^{3+}$- and the $Cr^{3+}$-spins contribute to the ESR signal. For temperatures above 10 K, the resonance field resides at about $H_{res} = 3.38$ kOe, but strongly shifts to lower fields at lower temperatures (Fig. 4b). Inserting the resonance field obtained above 10 K into the Larmor condition for magnetic resonance, $h\nu = g\,\mu_B\,H_{res}$, where $h$ denotes the Planck constant and $\mu_B$ the Bohr magneton. We obtain a $g$-value of 1.976, close to $g = 2$, expected for $Gd^{3+}$ ion with half-filled $4f$ shell and spin-only $Cr^{3+}$ ($3d^3$) with quenched orbital momentum. The resonance linewidth $\Delta H$ amounts to about 3 kOe at elevated temperatures, reveals a flat, broad maximum around 40 K and strongly broadens below approximately 10 K (Fig. 4c). The linewidth at elevated temperatures is comparable to that observed in $GdMnO_3$[20] and is a consequence of anisotropic exchange and dipolar interactions within and between the Gd and Cr spin systems.

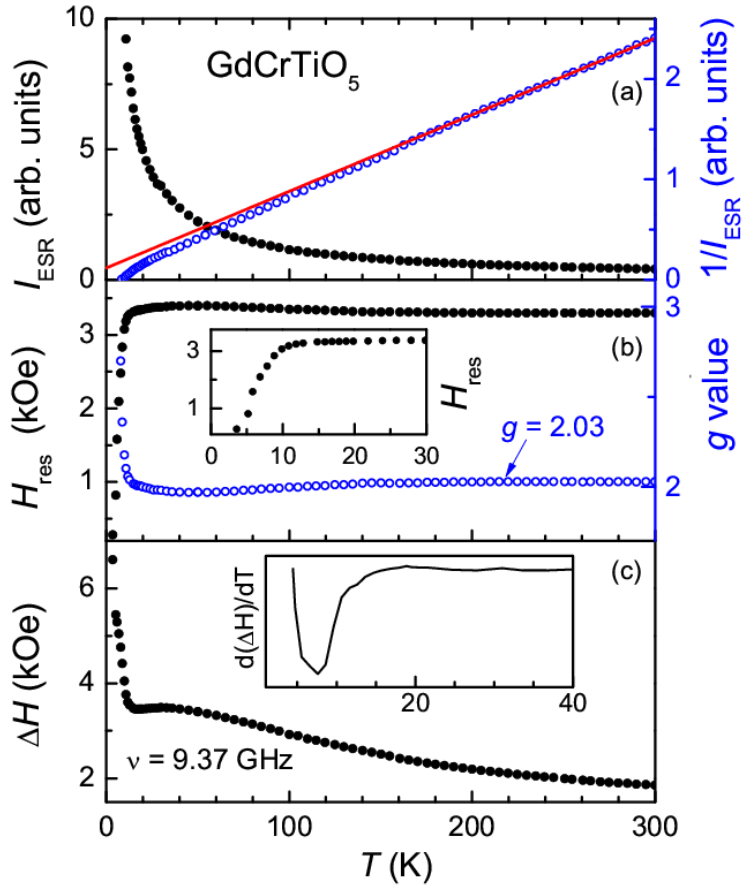

**Figure 4:** (a) The double-integrated intensity $I_{ESR}$ (left axis) and inverse intensity $1/I_{ESR}$ (right axis) are plotted as a function of temperature $T$ for the compound $GdCrTiO_5$. The red solid line shows the Curie-Weiss fit as described in the text. (b) Resonance field $H_{res}$ (left axis) and Landé $g$ factor (right axis) as a function of $T$. The inset highlights the $H_{res}$ data at low

temperatures. (c) Linewidth $\Delta H$ (HWHM) as a function of temperature. The inset shows the temperature derivative of $\Delta H$, to highlight the strong anomalies at low temperatures.

Both the strong and abrupt increase of the resonance field and the significant line broadening below 10 K (see Figs. 4b and c) clearly indicate the onset of some kind of magnetic order close to 10 K, most probably of the Cr moments only, while the Gd moments remain paramagnetic down to the lowest temperatures (Fig. 4a). The slight kink in the inverse ESR intensity close to 10 K (Fig. 4a) can also be interpreted along these lines. This type of behavior of the ESR line width and the resonance field has been observed in a variety of compounds revealing strong electronic correlations, which are close to magnetic order.[21,22] The broad hump or saturation effects, which are visible in the temperature dependence of the line width below 50 K (Fig. 4c), obviously correspond to deviations from the ideal Curie-Weiss behavior of the static susceptibility (Fig. 4a) due to short-range antiferromagnetic correlations.

## C. Zero-field μSR spectroscopy

To further explore the nature of the magnetic ground state in $GdCrTiO_5$, we performed zero-field μSR measurements, which again are very sensitive to changes of the local magnetic fields. The asymmetry $G_z(t)$ of ZF μSR spectra is shown as function of time $t$ at different temperatures below 10 K in Fig. 5. For a detailed analysis, the asymmetry of the muon decay is modeled with an exponential decay function ($G(t) = A_{bg} + A_0 \times e^{-\lambda t}$), where the constants $A_{bg}$, $A_0$ and $\lambda$, correspond to the background, the pre-exponential factor (i.e. the initial asymmetry) and the relaxation rate, respectively. Solid lines in Fig. 5 indicate the result of these fits. In the inset, the asymmetry as measured at 20 K is compared with that measured at 100 K indicating that only minor changes of the asymmetry of the μSR signal can be observed at elevated temperatures (Fig.5). Furthermore, the relaxation rates are very similar at 20 K and 100 K.

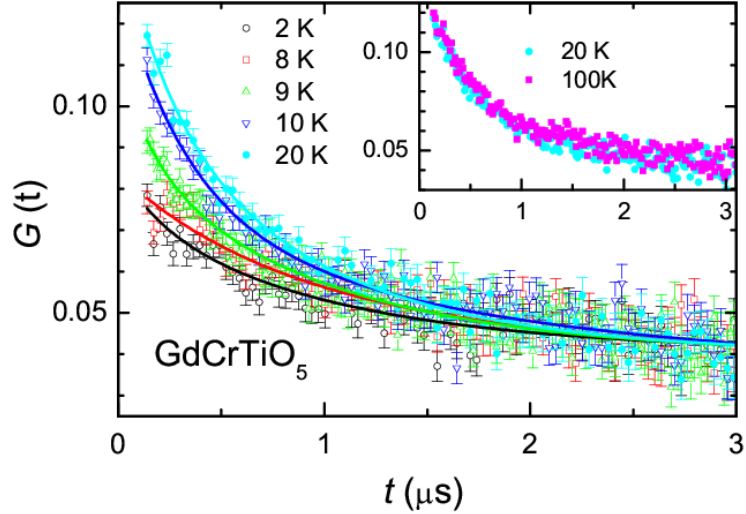

**Figure 5:** Asymmetry ZF μSR spectra observed in GdCrTiO$_5$ at different temperatures *vs.* time below 20 K. The solid lines represent fits as explained in the text. In the inset spectra taken at 20 K are compared with those taken at 100 K.

In Fig. 6 the resulting relaxation parameters, namely the pre-exponential factor $A_0$ and the relaxation rate λ, are plotted as a function of temperature on semi-logarithmic scales. The initial amplitude of the asymmetry spectrum (G(t) at t = 0) is nearly constant above approximately 11 K, exhibits a sharp jump around 10 K and becomes nearly constant below 8 K, as shown in Fig. 6a. Interestingly, the initial low-temperature asymmetry, after taking out of the background component, is nearly 1/3 of that at elevated temperatures (Fig. 6a). The 1/3 drop in the initial asymmetry is generally observed below long-range magnetic ordering temperature with a larger magnetic moment or larger internal field at the muons stopping sites. However, we did not observe any oscillations of the asymmetry spectra (Fig. 5), a typical signature of long-range ordered magnetic moments, as observed in many other magnetic compounds (and multiferroic as well).[23] When the internal fields are larger and with larger relaxation rate the signal damps faster and it could be difficult to observe any frequency oscillations using the μSR spectrometer at ISIS due to the pulse width.

Fig. 6b shows that the relaxation rate is constant at elevated temperatures, exhibits a jump-like reduction at 10 K and remains constant again towards low temperatures. Clearly, the striking anomalies in both parameters, asymmetry and relaxation rate, indicate significant changes of local magnetic fields. The results as documented in Fig. 6 strongly point towards a long-range magnetic order or spin-glass behavior in GdCrTiO$_5$ for temperatures below 10 K. In the present compound, only the chromium moments ($S = 3/2$) will order, while the gadolinium moments ($S = 7/2$) remain paramagnetic, as indicated[15] by the bulk susceptibility. Hence, below 10 K in GdCrTiO$_5$ two spin species exist, one, which is ordered, and one that still fluctuates. The ordered one could be screened by the fluctuating component. It is also possible that below 10 K due to the pulse width of muons produced at ISIS, we are unable to estimate both components from the present data. For example, the multiferroic compound HoMnO$_3$,[24] where Mn and Ho moments order at significantly different temperatures, did not reveal clear oscillations in the asymmetry spectra at the onset of magnetic ordering, when measured at the pulsed muon source at ISIS. However, heavily damped oscillations resulting from antiferromagnetic long-range order were captured from measurements performed with the continuous muon source at PSI (Switzerland).[24] We believe that the abrupt, jump-like decrease of λ(T) in GdCrTiO$_5$ at 10 K, probably signals a magnetic transition. However, the observed temperature dependence of λ(T) is not characteristic of that usually observed in magnetically long-range ordered systems. Here, one expects a rather continuous increase of λ(T) with lowering temperature above the long-range magnetic ordering and a rapid decrease below the magnetic ordering temperature (e.g., Ref. 24). Therefore, our µSR results with respect to temperature dependence of the relaxation rate and absence of oscillation in ZF-spectra strictly do not prove long-range magnetic order in GdCrTiO$_5$, even though the drop in the asymmetry by nearly a factor of 2/3 clearly points towards a magnetic ground state.

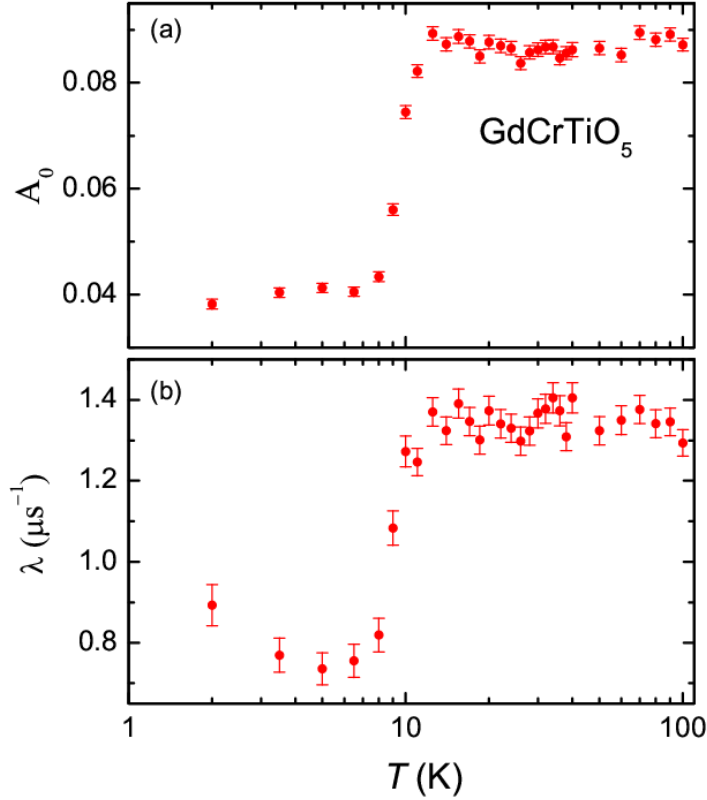

**Figure 6:** Temperature dependencies of the fitting parameters of the exponential decay of ZF µSR spectra in $GdCrTiO_5$ as discussed in the text. The data are presented on a semi-logarithmic plot: (a) initial asymmetry $A_0$ and (b) relaxation rate $\lambda$.

### D.  Spin driven ferroelectricity and magnetoelectric effects

The pyroelectric current as a function of temperature, with positive and negative electric poling fields, for various external magnetic fields from 0 to 80 kOe is shown in Fig. 7a. The data were recorded with a heating rate of 2 K/min. The resulting remnant polarization for positive poling fields is depicted in Fig. 7b for the same set of external fields. No pyrocurrent and hence no polarization is observed for external magnetic fields up to 10 kOe. A finite ferroelectric polarization appears below ~ 10 K only in the presence of a magnetic field $\geq 20$ kOe and only if the sample was cooled in an external magnetic field. The ferroelectric ordering temperature slightly increases by about 1 K for higher magnetic fields (see Figs. 7a and 7b). Higher magnetic fields also enhance the saturation polarization almost continuously up to 0.4 µC/m² for 80 kOe (Fig. 7b). Thus a magnetic-field induced ferroelectric phase can be inferred, which is depicted in Fig. 7c. The polarization can be switched for reversed electrical poling fields (Fig. 7a), confirming the ferroelectric behaviour. The material is insulating ($\tan\delta \leq 0.0007$ for $T < 25$ K) and we did not observe any dielectric relaxation in this system at the characteristic temperature (see Ref. 15). In addition, we reproduced the pyrocurrent behavior with a heating rate of 5 K/min (not shown here); no shift of the peak temperature is detected for different heating rates. These results altogether clearly exclude extrinsic artefacts to be responsible for the observation of a pyrocurrent signal, such as, thermally stimulated depolarization current or pyro-current due to hopping conductivity[25] and confirms the intrinsic ferroelectricity.

Interestingly, a finite value of the pyroelectric current is observed for FC conditions only. If the sample is cooled in zero external magnetic field, followed by measurements with applied magnetic fields, no pyroelectric signal can be observed within the detection limit. The blue cross symbol in Fig. 7a document this effect, showing the results of ZFC measurement with the measurement of pyrocurrent with subsequent application of 50 kOe. It seems that an alignment of the paramagnetic Gd moments for $T < 10$ K in an applied external magnetic field of $H > 10$ kOe enforces a canting of the Cr moments, which persists only in the presence of an external magnetic field $H > 10$ kOe. The presence of canted Cr-moments allows breaking the inversion symmetry and gives rise to the observed improper spin-driven ferroelectric effect.

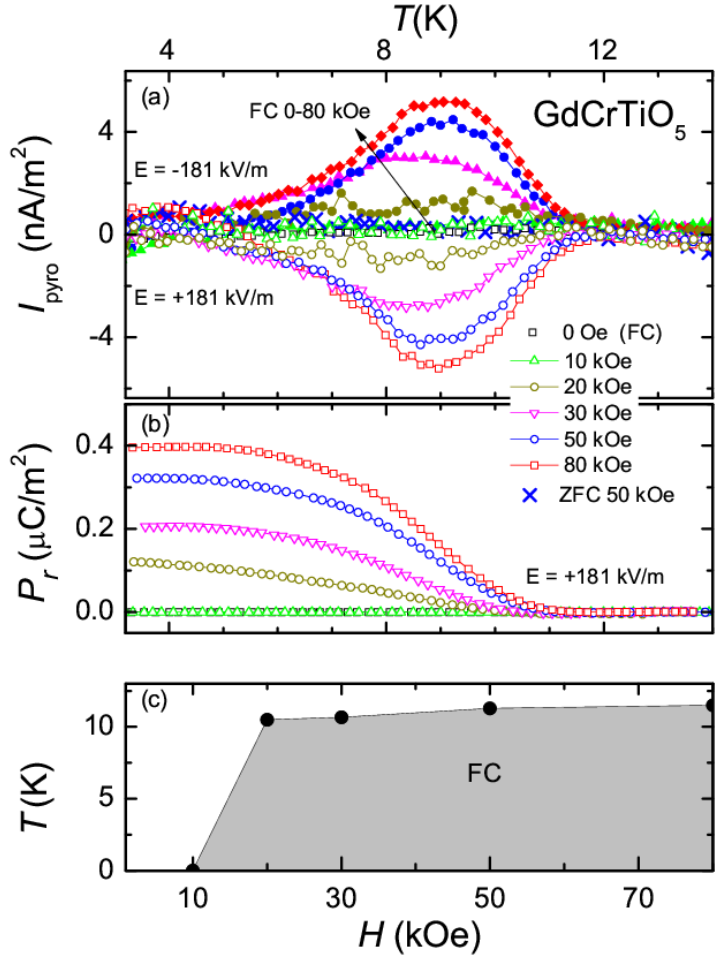

**Figure 7:** (a) Pyroelectric current of $GdCrTiO_5$ as a function of temperature for positive (open symbols) and negative (solid symbols) electric poling fields for different magnetic fields from 0-80 kOe in FC conditions as described in the text. For the 0 Oe and 10 kOe FC measurements the pyroelectric current is zero, therefore only data for positive poling fields are shown for clarity reasons. A measurement at 50 kOe taken under ZFC conditions (× symbol) is shown for comparison. (b) Remnant polarization for positive electric poling fields (E = 181 kV/m) for different external magnetic fields from 0-80 kOe as defined in (a). (c) ($T,H$) polar phase diagram of $GdCrTiO_5$. The shaded area indicates the ferroelectric phase.

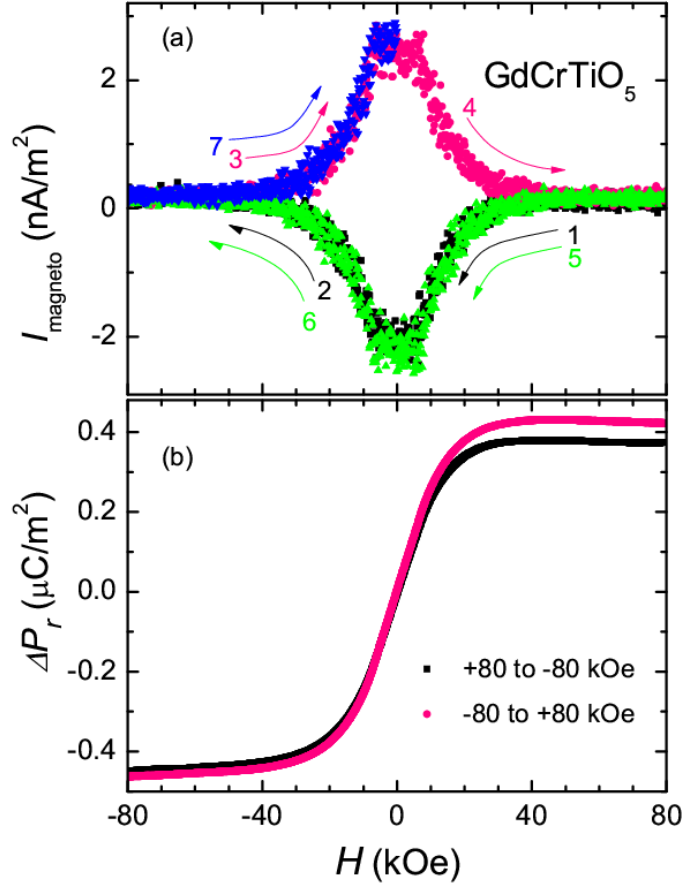

**Figure 8:** (a) Magneto-electric current for GdCrTiO$_5$ as function of magnetic field from +80 kOe to -80 kOe in FC condition at $T$ = 4 K. Black, red, green and blue curves (numbers 1-7) represent the magnetocurrent for different cycling directions of the magnetic field, as discussed in the text. $I_{magneto}$ is set to zero at the beginning of the measurement, i.e. at +80 kOe and therefore the values on y-axis are relative changes with varying $H$. (b) Relative change in remnant polarization as function of magnetic field at 4 K. The remnant polarization, $\Delta P_r$ is calculated as discussed in the text. Black and red symbols denote the two different magnetic field directions as depicted in (a).

Magnetocurrent measurements were performed at 4 K (Fig. 8a), which provide further experimental evidence for magnetoelectric effects in this compound. No change in the magnetocurrent is detected as a function of magnetic field, if the sample is cooled in zero-magnetic field, which is consistent with the $T$-dependent polarization. Therefore, prior to magnetocurrent measurement, the sample was cooled to 4 K in FC condition with +80 kOe. Thus, the measurement was started in the induced ferroelectric phase (i.e., +80 kOe at 4 K) and the current was detected as a function of the external magnetic field $H$ with continuous field sweeping. The change in remnant polarization $P_r$ ($H$) is calculated by integrating the current as function of time. This is a special case where the initial state is FC +80 kOe at t = 0s, unlike that reported for any other magnetoelectric compound until now. Therefore, the change in magnetocurrent and thus the change in polarization with $H$ are reliable, though absolute values may differ depending on measurement conditions. The relative change in polarization $\Delta P_r$ [= $P_r(H)$ - $P_r(H = 0)$] as function of magnetic field is plotted in Fig. 8b. The arrows in figure 8a denote the sequence of the magnetocurrent measurement. Reducing the magnetic field below about +40 kOe results in a decreasing magnetocurrent (Fig. 8a), which originates from a reduced polarisation based on a less strong canting of the Cr moments. With further decreasing magnetic field, at about +10 kOe, negligible change of magnetocurrent is observed (see sequence '1' in Fig. 8a). Increasing now the magnetic field value in the opposite direction up to -80 kOe gives rise to a symmetric magnetocurrent behaviour (sequence '2' in Fig. 8a). The electric polarisation switches into the opposite direction by reversing the magnetic field (c.f. Fig. 8b). Switching back into the initial state (sequence '3' and '4' in Fig. 8a) gives rise to a magnetocurrent with the opposite sign. Again, negligible change in magnetocurrent is detected between around -10 kOe and +10 kOe denoting the dielectric regime. The electric polarization curve (Fig. 8b) for the cycle -80 to +80 kOe (cycle '3' and '4') derived from the above mentioned magnetocurrent data almost overlaps with that for +80 to -80 kOe (cycle '1' and '2'), without showing any clear hysteresis (a small change in the polarization value is an artefact of the measurement). The sequence '5' to '7' (Fig. 8a) superimposes with sequence '1' to '3' respectively, which proves the intrinsic nature of the observed closed loop of the magnetocurrent signal while changing the magnetic field from +80 kOe to -80 kOe. It seems plausible that, for FC conditions (with Gd moments resulting in canting of Cr moments), a critical magnetic field ($H$ > |10 kOe|) exists, which breaks the inversion symmetry allowing spin-driven ferroelectricity. The fluctuations of Gd moments may hamper a stable arrangement of Cr spins, however, the FC condition could help to stabilize the spin texture of Cr moments by reducing fluctuation of Gd moments.

4.     **Discussion and Summary**

Unlike the isostructural neodymium compound, *dc* and *ac* magnetic susceptibilities as well as the heat capacity do not yield any signature of long-range magnetic order in $GdCrTiO_5$. However, ESR and µSR investigations document significant anomalies indicating clear changes of local magnetic properties in this system. These anomalies appear close to 10 K, the temperature where the onset of finite ferroelectric polarization can be detected in external magnetic fields > 10 kOe. One explanation for this experimental observation via local spectroscopic probes could be the onset of magnetic order of the chromium moments at 10 K. This magnetic order of the chromium spins with $S = 3/2$ is detected via microscopic probes, but remains hidden in the bulk susceptibility as well as in the thermodynamic response, due to the large fluctuating paramagnetic moment of the Gd ions with $S = 7/2$.

Compared to NdCrTiO$_5$, the magnetism of the Gd compound investigated in the course of this work seems to be strongly affected by the heavy rare-earth moment. In the neodymium compound, the chromium moments exhibit antiferromagnetic (AFM) order below 21 K, while the neodymium moments gradually order below approximately 13 K.[13] The chromium moments are aligned parallel to the crystallographic $c$ axis, with AFM order within the $ab$ plane and AFM stacking along the $c$ axis. The neodymium moments exhibit AFM order with the magnetic moments aligned within the $ab$ plane. In contrast, in the Gd compound, the magnetic order of the chromium moment is shifted to 10 K, while the Gd spins seem to fluctuate down to the lowest temperatures. If an external magnetic field is applied, a field induced quasi-ferromagnetic alignment of the paramagnetic Gd moments influences the Cr sublattice. It is clear, that all exchange interactions of the 4$f$ moments in this class of compounds are very weak and that the 4$f$ moments have to align in the internal magnetic field of the 3$d$ spins, which exhibit antiferromagnetic order. In the Gd compound the 4$f$ exchange is further weakened due to the spin-only character of the half-filled 4$f$ shell of the Gd ions, with negligible spin-lattice interaction. Hence, the Gd spins remain almost paramagnetic down to the lowest temperatures, even in the internal field of the ordered chromium moments. It may be also possible that stronger frustration of the interaction between competing Gd and Cr spins plays a role for lowering the ordering temperature in GdCrTiO$_5$ with respect to NdCrTiO$_5$: Comparing the susceptibilities of the Nd and Gd compounds, the value for the former is about a factor of 10 smaller at 20 K than in the latter.[13,15] This difference results from the large contribution of the Gd spins. In NdCrTiO$_5$ the Nd moments are forced to align antiferromagnetically within the $ab$ plane between the antiferromagnetic Cr layers with their magnetic moments along the $c$ axis. Obviously, thermal fluctuations of the larger Gd moments partially suppress the magnetic order of the Cr moments in this compound and, therefore, significantly decrease the ordering temperature compared to that for Nd member.

The spin-driven ferroelectric behaviour of GdCrTiO$_5$ is closely related to the observations in the Nd member of this series, as reported by Hwang et al.[13] However, there are clear differences in the field-dependent polarization. In the case of NdCrTiO$_5$, there is a linear increase of polarization, starting from zero magnetic field,[13] eventually ferroelectricity is reported in absence of magnetic field by Saha, *et al.*[26], whereas in GdCrTiO$_5$, finite values of the pyroelectric current are observed only in the presence of external magnetic fields > 10 kOe. This is documented by magnetic field-dependent pyrocurrent measurements (i.e. magnetocurrent) performed in the course of this work. The ferroelectric ordering temperature slightly increases with increasing magnetic fields for GdCrTiO$_5$, whereas, magnetic field effects seem less important for the ferroelectric ordering temperature of NdCrTiO$_5$.[13,26]

An interesting observation of the present work is the drop in ferroelectric polarization below 10 kOe in the magnetocurrent measurements for the material cooled in FC condition below the ferroelectric ordering temperature. Surprisingly, no metamagnetic-like transition is observed in magnetization experiments. However, the weak metamagnetic transition of the Cr system is probably masked by the strong paramagnetic contributions of the Gd moments, whereas, the paramagnetism of Gd may not affect the electric features. It appears, that ME measurements sometimes are more sensitive to detect such field-induced transitions, compared to magnetic measurements, at least in this case. Here, GdCrTiO$_5$ exhibits strong ME coupling and significant effects of external magnetic fields on the ferroelectric behaviour; however, there is no direct one-to-one correspondence of observed anomalies in the H-dependent isothermal magnetic and electric behaviour. This kind of unusual ME behaviour due to complex dynamics of spins and dipoles is rare in the literature; for example, a magnetodielectric phase-coexistence is observed in the spin-chain compound Ca$_3$Co$_2$O$_6$ without any fingerprints of such a phase-co-existence in isothermal magnetization measurements, despite strong intrinsic magnetodielectric coupling in this complex compound.[27]

The results suggest that spin-induced ferroelectricity in *R*CrTiO$_5$ may be governed by DM interactions due to the Cr moments, significantly influenced by *R*-moments. It has to be noted that a finite magnetodielectric coupling (dielectric constant as a function of *H*, which consists higher order coupling term as well) for GdCrTiO$_5$ exists even for temperatures up to 20 K.[15] This fact could arise from magnetic exchange frustration resulting from competing exchange interactions between Gd and Cr moments, active even above magnetic ordering (> 10 K). For GdCrTiO$_5$ we speculate about a magnetic field induced quasi-ferromagnetic alignment of Gd moments for *T* < 10 K enforcing a canting of Cr moments, again in the presence of a magnetic field *H* > 10 kOe. This canting of Cr moments allows inverse DM interaction that breaks the inversion symmetry leading to spin-driven ferroelectric ordering. Note that, only if the system is "frozen" in a frustrated state via cooling below 10 K with applied extrinsic magnetic field > 10 kOe, the Cr moments form a stable canted spin structure in an applied magnetic field and break the inversion symmetry which govern ferroelectricity.

While our manuscript was under review, a manuscript has been published by Guedes *et al.*[28] reporting on the magnetism of $GdCrTiO_5$. They demonstrated the onset of long-range magnetic order of the Gd spins at 0.9 K and suggested non-collinear antiferromagnetic ordering in this system. This recent report[28] supports our interpretation of the magnetic properties of $GdCrTiO_5$. Hence, our study confirms that dielectric and pyroelectric measurements are very sensitive to capture such weak or metastable *H*-induced transitions, especially in the case of strong ME coupling.

In summary, the magnetodielectric compound $GdCrTiO_5$ has been investigated by electric polarization measurements, ESR and ZF-μSR spectroscopy. Our results confirm significant changes of magnetic interaction below ~10 K, attributed to the onset of a possible canted magnetic order of the chromium moments. The role of the rare-earth ions on the magnetic and magnetoelectric features is also inferred. The ferroelectric polarization is observed only for external magnetic fields > 10 kOe and in field-cooled conditions, which is rare in literature.

**Acknowledgment**

This work was supported by the BMBF via project ENREKON 03EK3015 and partly by the DFG via the Transregional Collaborative Research Center TRR 80 (Augsburg, Munich, Stuttgart).